\documentclass[aip,reprint,twocolumn,groupedaddress]{revtex4-1}
\usepackage{color}
\usepackage{graphicx}
\usepackage{amsmath}
\usepackage{latexsym}
\usepackage{graphicx}
\usepackage{grffile}
\usepackage{amssymb}
\usepackage{amsfonts}
\usepackage[margin=1.in]{geometry}
\usepackage[autostyle]{csquotes}

\begin{document}

\title{Nonlinear resonances and multi-stability in simple neural circuits}

\author{Leandro M. Alonso}
%\email{lalonso@rockefeller.edu}
%\email{\\ lalonso@rockefeller.edu \\ leandro.alonso.ruiz@gmail.com}
%\affiliation{The Rockefeller University, New York, NY 10065, USA.}
\email{leandro.alonso.ruiz@gmail.com}
\date{\today}
\begin{abstract}
This article describes a numerical procedure designed to tune the parameters of periodically-driven dynamical systems to a state in which they exhibit rich dynamical behavior. This is achieved by maximizing the diversity of subharmonic solutions available to the system within a range of the parameters that define the driving. The procedure is applied to a problem of interest in computational neuroscience: a circuit composed of two interacting populations of neurons under external periodic forcing. Depending on the parameters that define the circuit, such as the weights of the connections between the populations, the response of the circuit to the driving can be strikingly rich and diverse. The procedure is employed to find circuits that, when driven by external input, exhibit multiple stable patterns of periodic activity organized in complex tuning diagrams and signatures of low dimensional chaos.  
\end{abstract}
\maketitle

\begin{quotation}
This article introduces a numerical procedure designed to tune dynamical models of interacting populations of neurons towards states in which an incoming periodic signal may trigger a variety of qualitatively different responses. Even in the simple case of two interacting populations, there are specific connectivities for which the circuit can entrain nonlinearly with an incoming signal. The procedure consists of optimizing an objective function that measures the diversity of responses by counting how many different subharmonic resonances are within the range of the incoming signal. When driven by periodic input, the resulting circuits can present a multitude of stable oscillatory responses organized by complex locking diagrams. The procedure presented here may be useful to further elucidate the role of chaos and nonlinear oscillations in biological phenomena.
\end{quotation}

\section{Introduction} 
The brain is a multi-rhythmic system with spatiotemporal activity spanning several orders of magnitude which range from fast localized spiking at the single unit level to macroscopic oscillations that involve many units firing in synchronous patterns. Neural oscillations are believed to underlie a wide spectrum of brain functions (see Buszaki et al. for a review \cite{buszakiscience}). The coordination of motor patterns utilized in animal locomotion and maintenance processes such as breathing are controlled in part by central pattern generators \cite{golubitsky,delcomyn}. There is a growing view that rhythmic neural activity plays an active role in shaping neural processing and behavior by transiently binding cells into synchronized assemblies \cite{singernature}. Different rhythms may shape the effective connectivities of local circuits and alter the way information is processed and routed. Recently, experiments in rodents have shown that theta (8 $Hz$ to 9 $Hz$) oscillations in the rats hippocampus encode the animals position in an arena \cite{buzsakilfp}. The emerging view is that neuronal oscillators may provide a basis for a variety of neural functions including cognition \cite{buszakilibro}.

Nonlinear resonances in periodically-driven neural circuits have previously been studied both theoretically and experimentally in several contexts. Thalamo-cortical interactions were studied using models of weakly connected oscillators in which communication between cortical columns is enabled by resonances \cite{izifm}. Additional examples in the neuroscience literature include the amplification of gamma rhythms in inhibition-stabilized networks \cite{sejnowski}. Recently, studies of periodic stimulation in models of spiking neurons have shown that intrinsic network oscillations can be shaped by changing the amplitude and frequency of the stimulation, suggesting possible implications for trans-cranial stimulation methods \cite{lefebvrejon}. Periodic stimulation of nonlinear physical and biological systems constitutes a traditional approach to characterizing their intrinsic dynamical properties. This approach was used to investigate the intrinsic dynamics of a small network of electrically coupled neurons in the pyloric central pattern generator of the lobster \cite{abarbanelpyloric}. More recently, it was shown that periodic stimulation of telencephalic nuclei in songbirds creates subharmonic entrainments of the respiratory network \cite{mendez}. 

The notion that chaotic dynamics might underlie neural dynamics has been explored by many authors. Two aspects of chaotic dynamics that are argued to be relevant for brain function are sparse exploration of the phase space and built-in multipurpose flexibility. Even though the dimension of a chaotic attractor is in general smaller than the allowed state space, chaotic motions explore a broad region of the behaviors available to the system, providing a mean to seize opportunities upon environmental changes \cite{rabinovichroleofchaos}. Neural systems are required for different purposes at different times and under different conditions. A chaotic system can accommodate this type of multipurpose flexibility by switching the temporal programming of a small parametric perturbation in order to stabilize the different orbits embedded within the attractor \cite{ottcontrol}. More recently, Mindlin et al. explored the possibility that the diversity of respiratory motor gestures in birdsong can be partly explained by nonlinear entrainments and bifurcations of a driven hierarchy of neural nuclei  \cite{prlsubarmonicos,granada,pranama,alonso09,goldinmindlin,gogui}.

This article introduces a numerical procedure designed to tune dynamical models of neural activity toward special regimes in which many qualitatively different behaviors are available. The procedure is applied to the case of simple circuits composed of interacting subpopulations of neurons which receive oscillatory input from elsewhere. The response of a circuit to a given stimuli will vary depending on the parameters which define the architecture. Additionally, the response of a given circuit may be quite different as the frequency or the amplitude of the incoming stimuli change. Here we are interested in a hypothetical scenario in which the response of the circuit exhibits maximal diversity when receiving external stimuli within a given range. This scenario can be attained by asking that the circuits can be entrained to many different nonlinear resonances. The procedure specifically consists of computing a low resolution approximation of the Arnold tongues diagram of the circuit over a fixed range of stimuli parameters. Each point in the range is assigned an integer corresponding to the period of the response (if it is periodic) in units of the stimuli period. The resulting diagrams are scored by counting how many subharmonic solutions occur in the inspected range and asking that there is an equal number of solutions for each subharmonic type. This approach yields simple circuits that exhibit patterns of periodic activity with multiple timescales and diverse waveforms, as well as complex non-periodic behavior. The procedure can be used to further explore links between chaos theory and neural oscillations. 

This work is organized as follows. A numerical procedure to generate circuits which respond sub-harmonically to a family of periodic stimuli is presented in Sections 2a and 2b. The procedure is applied in a simple setting of two interacting populations of neurons described in Section 2c. Several examples of such circuits are presented and discussed in Section 3. Section 4 contains final remarks and future directions. 

\section{Methods} 

\subsection{Definition of locking period. Description of the algorithm.}

This section introduces a numerical procedure designed to tune the parameters of periodically-driven dynamical systems towards regimes in which several subharmonic entrainment regimes are possible. Let $F(x,p) : \mathbb{R}^{m \times q} \rightarrow \mathbb{R}^m$ be a smooth vector field with state variables $x \in \mathbb{R}^m $ and parameters $p \in \mathbb{R}^q$. It is further assumed that the system is driven by a family of periodic signals of period $\tau$, $\gamma_{\alpha}(t)$ parameterized by $\alpha$, 
\begin{eqnarray} 
\dot{x} &=& F(x,p + \gamma_{\alpha}(t))           \nonumber \\
x(0) &=& x_0.
\label{ode}
\end{eqnarray}

The purpose of this procedure is to find parameters $p$ such that the system will exhibit subharmonic entrainments with the driving signal $\gamma$, ie: that there is an integer number of periods $p \geq 1$ such that
\begin{eqnarray} 
  x(t + k (p \tau)) = x(t)  \qquad \forall k \in \mathbb{Z} \quad \forall t \in \mathbb{R}. 
\end{eqnarray}
In order to test if a given solution is entrained to the driving signal, a Poincar\'e section of the flow is computed by annotating the state of the system $x$ stroboscopically using the frequency of the driving signal \cite{gucken,wiggins}. This is done by taking regularly spaced timestamps every period of the forcing $\tau$ for a maximum number of periods $M$. This yields a sequence of states $\{x_0, x_1, . . ., x_M\}$ which are used to compute the mismatch between the first state $x_0$ and successive states $x_n$, 

\begin{eqnarray} 
E_n =  \parallel x_n - x_0 \parallel^2 .
\end{eqnarray}
Finally, the locking period $l_p$ is defined as the lower $n$ such that $E_n < \epsilon$,
\begin{eqnarray} 
l_p = \min_n \{n \in [1,M]:  E_n < \epsilon\},
\end{eqnarray}
where $\epsilon$ is a parameter of the procedure corresponding to the numerical accuracy utilized to determine the mismatch between the state of the system in the control sections. If there is no $n$ that satisfies $E_n < \epsilon$, we assign the value $M+1$ indicating that the system does not entrain up to period $M$. 

In order to compute the Poincar\'e sections, the system is integrated using a Runge Kutta $O(4)$ method using $dt = \frac{\tau}{100}$ \cite{numericalrecipes}. Before the Poincar\'e section is computed, the system is allowed to relax to the attractor for a number of transient periods $M_t$. This provides a simple numerical way to check if the system is entrained, but it may provide the wrong answer if the transient decays to the attractors are too slow or if the numerical resolution parameter $\epsilon$ is set too high. An additional caveat is that a fixed point solution will be identified as a period $1$ locking. This ambiguity was found to be unimportant for the purposes of this work. This definition provides a map between the parameters of the driving signal $\alpha$ and an integer number corresponding the period of the resulting solution. It is employed here to roughly quantify the diversity of dynamical responses available to a given set of inputs. In the case of weakly interacting oscillators these diagrams have specific shapes in the space of frequencies and amplitudes which resemble a V. They are known as Arnold tongues after V.I. Arnold and they are the objects which inspired this procedure \cite{arnold}. 

\subsection{Objective function}

Next we introduce an objective function to drive the parameters of system (\ref{ode}) towards nontrivial entrainment regimes. In general, nonlinear systems will exhibit complex resonances when driven with periodic input. Small variations in the features of the input may result in different locking regimes which can be partially characterized by an integer. This maps the space of stimuli features into a set of integers which define the different entrainment regions. The set of external stimuli is specified by defining a domain for the stimuli parameters $\alpha$, and the locking period is computed in a grid of $N$ regularly spaced points in this domain. A discrete distribution of locking periods $\{L_j\}$ is built by counting how many solutions of each period were found over the chosen domain. The goal of this procedure is to obtain circuits that are close to many nonlinear resonances. One way to achieve this is to ask that the distribution of locking periods is flat, ie: that there is an equal number of solutions for each locking period (up to period $M$) within range. This is attained by minimizing an objective function, 
\begin{eqnarray} 
\label{costfun}
C(x_0,p) =  \sum^M_{j=1} (\frac{L_j}{N} - \frac{1}{M})^2.
\end{eqnarray}

Evaluation of the objective function requires determining the locking period of a number $N$ of solutions. This is the computationally intensive part of the procedure. To summarize, the computational effort required to evaluate the objective function (\ref{costfun}) scales linearly with the number of solutions to evaluate $N$, and linearly with the maximum number of periods $M$ and with the transient periods $M_t$. As discussed in the next section, there are choices of these values for which evaluation of this function is fast enough so that the problem can be tackled by several heuristic optimization approaches. 

\subsection{The model}
As a case study, this article considers dynamical models of neural activity which can be cast in the form of an ordinary differential equation or vector field. The procedure is applied to study the dynamics of neural populations using the celebrated Wilson-Cowan model \cite{wilsoncowan72}. This model and its extensions have been widely used to model neural populations and it is a common approach to address several problems in computational neuroscience \cite{destexhewilsoncowan}. Here we consider the simplest case of two populations of interconnected neurons of excitatory and inhibitory subtypes. The state variables are a measure of the activity of each population. In the absence of stimulus, this system can present limit cycles and fixed-point behavior. However, when driven by periodic input, the dynamics can be extremely complex depending on the precise weights of the connections and other parameters which define the architecture. Here we study a family of circuits given by 
\begin{eqnarray}
\label{model}
\frac{1}{\tau_1} \dot{x_1}  &=&  -x_1 + S(C_{11} x_1 + C_{12} x_{2} + \rho_1 +  \gamma(t) )  \\ \nonumber
\frac{1}{\tau_2}  \dot{x_2}  &=&  -x_2 + S(C_{21} x_1 + C_{22} x_{2} + \rho_2 ),    \\ \nonumber
\end{eqnarray}
where $S$ is the sigmoid function, 
\begin{equation}
  S(x)  = \frac{1}{1 + e^{-x}}.
\end{equation}.

The timescale of each population is controlled by parameters $\tau_i$. Each population receives input from the rest via the connectivities $C_{ij}$ and a constant input $\rho_i$. We assume that an external input $\gamma(t)$ is injected into population $1$ and we are interested in the possibility that the circuits will respond in qualitatively different ways when driven by similar stimuli. Alternatively, the incoming signal can be thought of as produced by another neural oscillator and therefore the full circuit would present multiple stable patterns of periodic activity which can be switched by changing the weights of the connectivities or the offsets activities. We assume a particular family of periodic signals $\gamma$ that models an external neural oscillation defined by 
\begin{eqnarray}
\label{forcing}
\gamma(t) = \rho + A S(\eta(\cos(\omega t)-\mu))).
\end{eqnarray}
Here $A$ is the amplitude, $\omega$ is the frequency and $\rho$ is a constant offset. Parameter $\rho$ is included in equation (\ref{forcing}) in order to study modulations both in the amplitude and frequency $(\omega, A)$ as well as modulations in the amplitude and offset $(\rho, A)$. Parameters $\eta =0.75$ and $\mu=-1$ are used to control the waveform of the oscillation and are kept fixed in this work. These parameters were chosen so that the forcing is close to a purely sinusoidal function. Because the focus of this article is to obtain circuits that respond differently to similar stimuli, these parameters were kept fixed for clarity but they can be included in the optimization procedure in the same way as the rest of the parameters. These parameters were kept fixed so that the forcing signal is the same for all the circuits discussed in this work. Therefore, differences in the dynamical properties of each circuit arise from different architectures and not from changes in the spectral content of the forcing signal $\gamma$.  

\section{Results}
In this section we discuss circuits defined by equations (\ref{model}) which were obtained by optimization of the objective function (\ref{costfun}). For visualization purposes, two scenarios are considered independently. In the first case, the driving signal $\gamma$ can be modulated in amplitude and frequency, $\alpha=(\omega,A)$. In the second case, the amplitude and constant offsets are allowed to vary, $\alpha=(\rho,A)$. The connectivities are allowed to take a wide range of values $C_{ij} \in [-20,20]$, $\rho_j \in [-20,20]$ and the timescales $\tau_i = 1$ are kept fixed for simplicity. In order to evaluate the cost function we specified a domain for the stimulus parameters $\alpha={ A \in [0,10] , \omega \in [1-0.2 , 1+0.2], \rho \in [-5,5]}$. For each scenario, we took a regular grid of $10 \times 10$ values in the corresponding domains. We set the maximum number of locking periods $M=10$ and the numerical resolution parameter $\epsilon =0.001$. The solutions were allowed to decay to the attractors for $M_t = 10$ periods. Evaluation of the cost function thus entails determining the locking period for $N = 100$ regularly spaced points in the corresponding $\alpha$ domain up to period $M = 10$. For these parameters and in our current implementation we achieve about $15$ evaluations per second in a commercially available linux server. This makes the problem tractable by many heuristic approaches which do not require knowledge of the target function. The objective function (\ref{costfun}) is optimized by a genetic algorithm starting from $50$ random seeds in the search domain which were evolved for $200$ generations \cite{holland}. Each of the circuits presented here were found in about $10$ minutes of computer work and their parameters are summarized in table I. 

Figure 1 shows the locking diagrams corresponding to several evaluations of the objective function (\ref{costfun}). The locking period is computed in a $10 \times 10$ grid and indicated in colors (color bar in Fig. 2). Values close to $1$ correspond to circuits which do not respond to the periodic stimuli or wherein the response is trivial, while lower values correspond to more colorful diagrams for which several different entrainments occur. The key realization is that the underlying structure of the locking regions is such that it can be detected by sampling only a few points. The system can accommodate locking regions which consist of well-defined, connected large \enquote{blobs} in the space of stimulus features $\alpha$. The problem of optimizing the objective function is computationally feasible because the structure of the locking diagrams can be partially characterized by sampling the space of stimuli features with low resolutions. The main result of this article is that optimization of this function is possible and yields a procedural mechanism to generate neural circuits with rich dynamical properties. 

\begin{figure*}
\includegraphics[width=140mm]{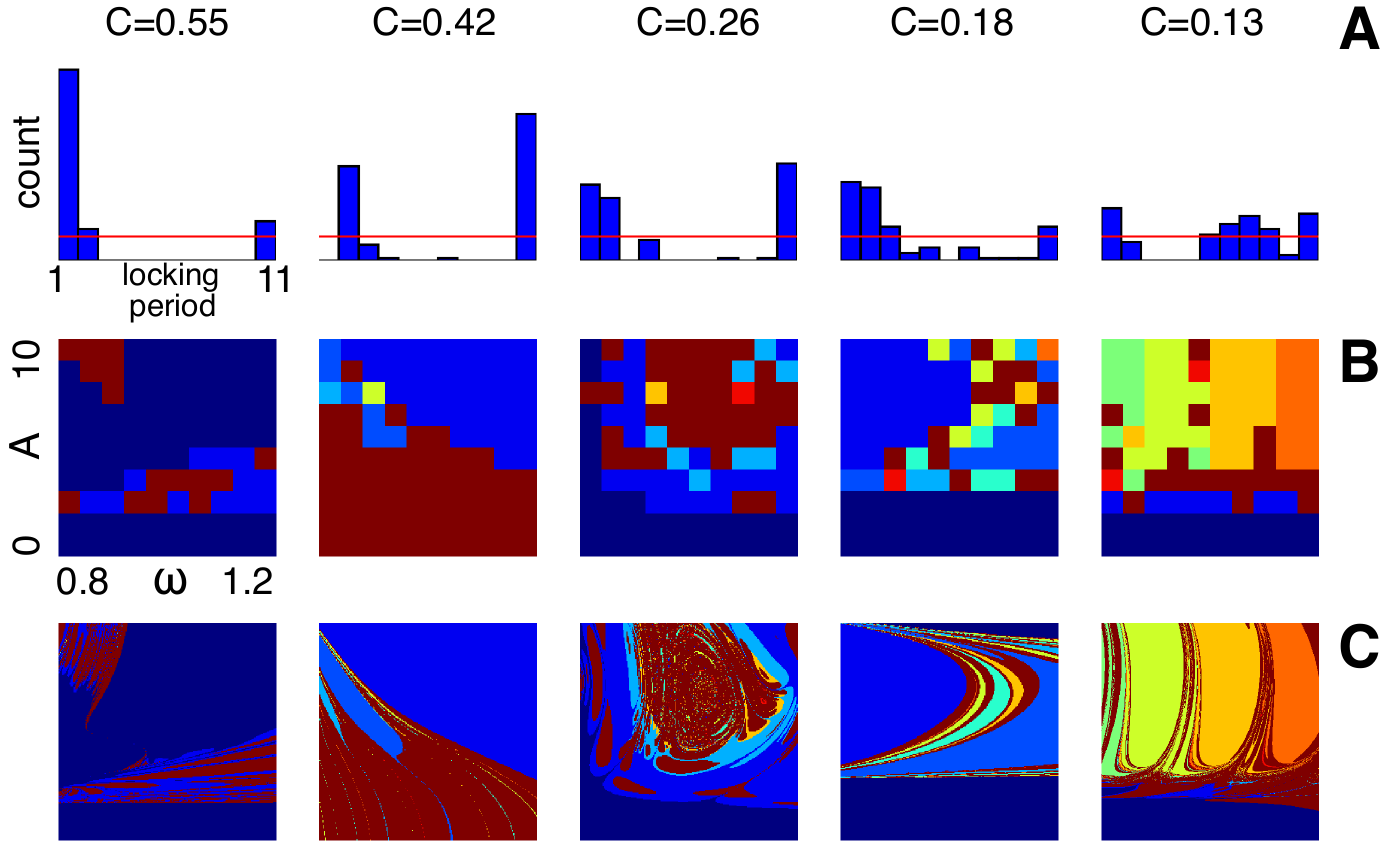}
\caption{\textbf{Evaluations of the objective function.} The circuits are obtained by minimizing the objective function. A single evaluation of the objective function consists of computing the locking period between the circuits response and the stimulus in a $10 \times 10$ fixed grid of stimulus parameters $(A,\omega)$ within a specified domain. This yields a map between stimulus parameters and integers, and a distribution of locking periods over the computed domain. The objective function measures how far this distribution is from being flat (the red line corresponds to a flat distribution). \textbf{(A)} Distribution of locking periods over the computed domain (range shown in figure). \textbf{(B)} Map between stimulus parameters and locking period. Locking periods are color coded ranging from period 1 (blue) to period higher than 10 or no locking (red). The color bar is shown in Fig. 2. \textbf{(C)} Increasing the resolution of (B) reveals the intricacies of the response diagrams.}
\end{figure*}

Figure 2 shows the response of a circuit as the amplitude and the frequency of the periodic forcing are allowed to change. We are interested in the case in which multiple timescales emerge out of the interaction of the populations. Figure 2A shows the response of the circuit for different values of the amplitude $A$. The system can be entrained to a multitude of oscillatory patterns with diverse temporal structure while the frequency of the stimuli remains fixed. Figure 2B corresponds to the locking period diagram over the inspected domain. Note that while the structure of the diagram is complex, there are large connected regions for each locking period. These regions are in turn separated by smaller nested structures of bands that exhibit beautiful patterns. In many cases, the waveform of the oscillations is different for each region. The dashed line in Figure 2B indicates the frequency of the stimulus used in 2A. This value was chosen so that several different regions became available by changing only the amplitude.  

The dynamics of the system in areas of the diagrams in-between the larger regions is often associated with strange attractors. Transitions from periodic to chaotic dynamics are known to exhibit universal scaling rules and there are specific routes by which chaotic attractors can develop in low dimensional systems \cite{feigenbaum,libchaber}. Figure 2C shows the dynamics of the circuit as the amplitude values are changed in small decrements towards a transition between regions ($A \in [2.705,2.740]$). The solutions are shown in full phase space $(x_1,x_2,\gamma)$ in order to visualize changes. As the amplitude is increased, the initial curve seems to split in two copies of itself which fail to intersect, in what appears to be a period doubling bifurcation. Further decrements of the amplitude result in successive folds of this mechanism until the rightmost attractor which is non-periodic and possibly chaotic. Period doubling bifurcations and possible chaotic behavior in neural models of the sort studied here, were previously reported by Ermentrout \cite{ermentroutchaos}. The underlying dynamical mechanisms by which these different oscillations emerge is beyond the scope this work as it depends on the specific details of each circuit. While in many cases the transitions from one region to another seem to correspond to local bifurcations, such as period doubling, there are other mechanisms by which complicated dynamics may emerge. Despite the complexity of the solutions there are a finite number of ways by which this can occur due to the low dimensionality of the system. The mechanisms underlying the generation of complex dynamics in low dimensional systems can be partly characterized and classified by topological analysis \cite{alicestretchland,classification}. 

\begin{figure*}
\includegraphics[width=140mm]{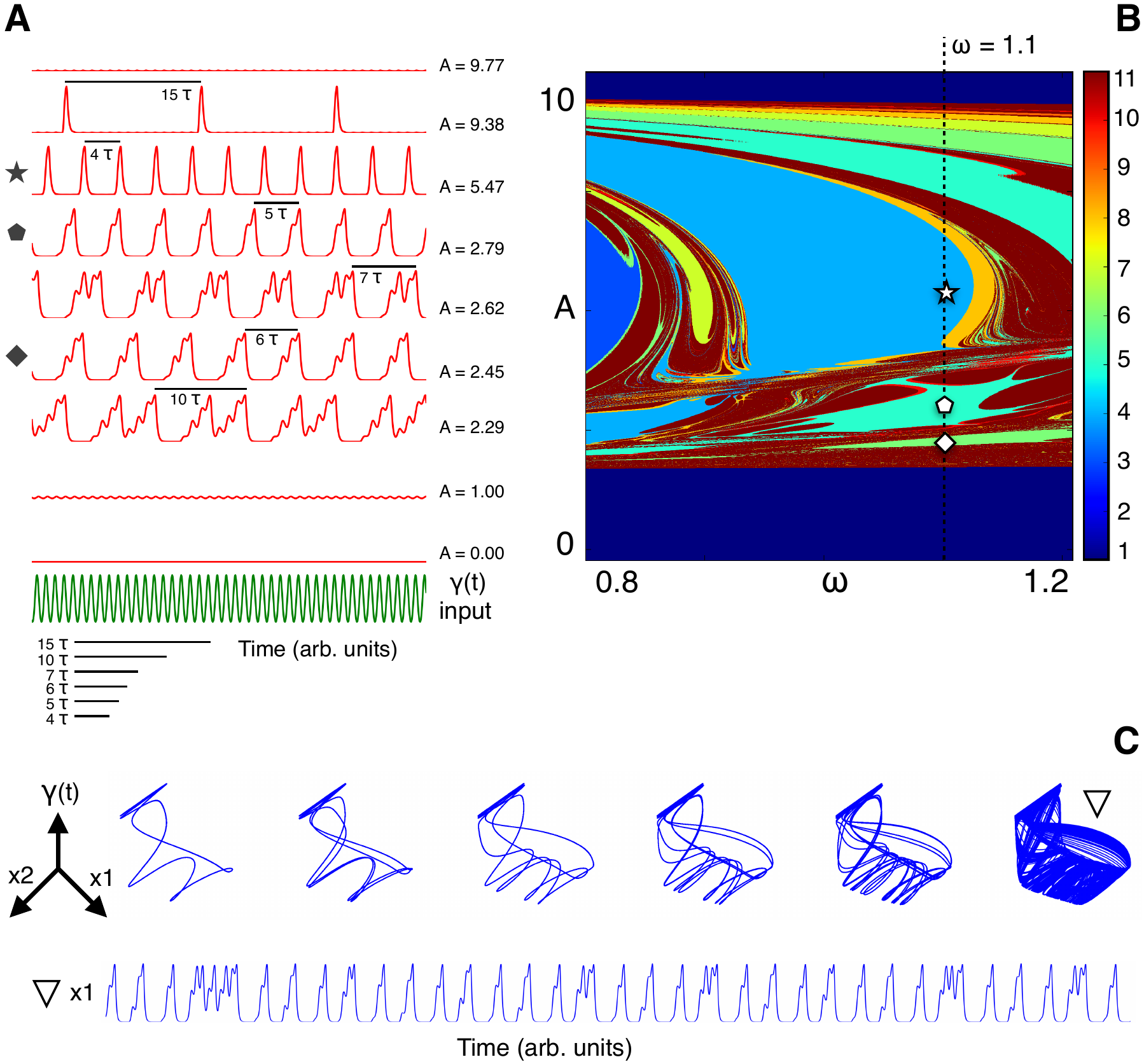}
\caption{\textbf{Emergence of multiple stable rhythms with complex organization.} When neural circuits are close to nonlinear resonances many dynamical behaviors become possible. The figure shows the response of a neural circuit consisting of two interacting populations of neurons $(x_1,x_2)$ receiving periodic input $\gamma(t)$, for increasing values of the amplitude of the input. \textbf{(A)} Responses of the circuit (red) as the amplitude of the stimulus (green) is increased. Both for low and high values of $A$ the system entrains 1 to 1 and the amplitude of the response is low. For intermediate values of $A$, the system displays a multiplicity of stable periodic rhythms with complex waveforms. \textbf{(B)} Locking period diagram for the considered domain of stimulus parameters $A$ and $\omega$. The locking period is color coded ranging from 1 (blue) to 11 (red, no locking or higher than 10). The dashed line indicates the frequency of the stimulus for all shown solutions. The symbols indicate solutions shown in (A). There are large connected regions for any considered locking period separated by nested structures of bands. This structures are in turn associated with complex dynamics shown in (C). \textbf{(C)} Simulations for decreasing values of $A$ near a transition between large regions plotted in phase space ($A \approx 2.705$). As the amplitude is decreased the solutions undergo period doubling bifurcations and possibly low-dimensional chaos.}
\end{figure*}

This procedure can also be applied to the potentially more interesting scenario in which the stimulus frequency is fixed and only the connectivities are changed. This results in circuits which can be slowly modulated to produce patterns on multiple timescales while being driven at a unique fixed frequency. We obtained circuits with this property by allowing $A$ and $\rho$ to change and optimizing the objective function (\ref{costfun}). Figure 3 shows the locking diagrams of one such circuit. The leftmost panel corresponds to the domain used in the optimization, while the successive panels zoom into the spiral. As before, the diagrams exhibit beautiful nested patterns which look self-similar. There are wide connected regions which are associated with a particular locking period, and typically, these regions are separated by thin bands that correspond to different periods. We can think of an extended system by including $\dot{A}=0 ,\dot{\omega}=0,\dot{\rho}=0$ and then the diagrams can be interpreted as different basins of attraction in the space of initial conditions $(A_0, \omega_0, \rho_0)$. From this point of view, the observed nested band structure is reminiscent of the Wada property for the basins of attraction of multi-stable nonlinear systems \cite{yorke}. 

\begin{figure*}
\includegraphics[width=170mm]{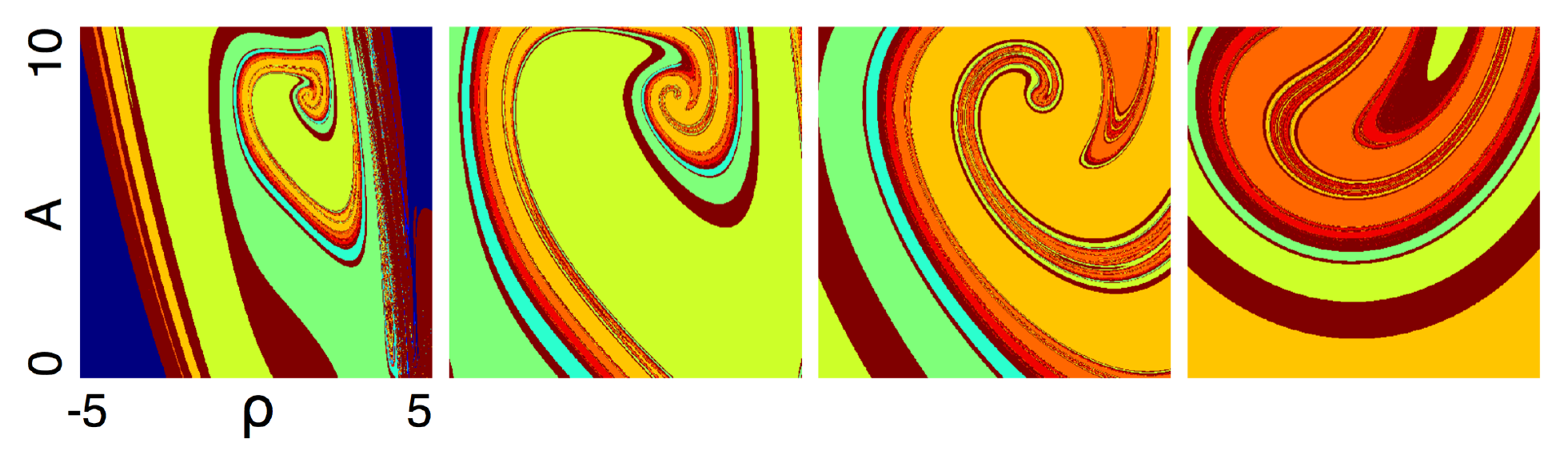}
\caption{\textbf{Locking period diagrams for modulations in amplitude and offset.} Circuits can also be tuned close to nonlinear resonances by considering modulations in other parameters. The diagrams show the locking period as the amplitude $A$ and the constant offset $\rho$ of the stimuli are allowed to change. There are large regions corresponding to the same locking period which are separated by complex structures of nested bands. This is highlighted in the zoomed in diagrams (b,c,d) to the right. The rightmost panel is zooming to the point $(A, \rho) = (8.010, 1.4965) \pm 0.001$.}
\end{figure*}

Finally, in order to allow for more biological realism in the resulting circuits, the optimization was performed in an extended domain including the possibility that the time scales of the populations are different. This additional degree of freedom results in lower values of the cost function and into the stunningly diverse locking diagrams shown in Figure 4. It is noteworthy that these type of models are able to accommodate such remarkable properties despite being relatively simple. This scenario might not be achievable in other systems for meaningful ranges of their parameters. Neural models of the sort studied here are remarkably flexible not only because they can be connected in many different ways, but also because the intrinsic properties of the nodes can be different. 

In the case considered here, the dynamics of the non-driven autonomous systems differ radically from the driven case. Because the circuits are composed of two populations, the dimension of the autonomous systems is $2$, implying that their solutions are either fixed points or limit cycles, for all values the parameters that define the architectures. As discussed in this article, this is not the case when the system receives periodic input. Since the dimension of the driven system is $3$, the system can in principle accommodate complex dynamical attractors, as those shown in Figure 2. The results presented here show that optimization of function (\ref{costfun}) constitutes an efficient procedure to achieve rich dynamical properties in circuits composed of two neural populations, and suggest that the procedure may be applicable to other systems of interacting nonlinear oscillators. 

\begin{figure*}
\includegraphics[width=160mm]{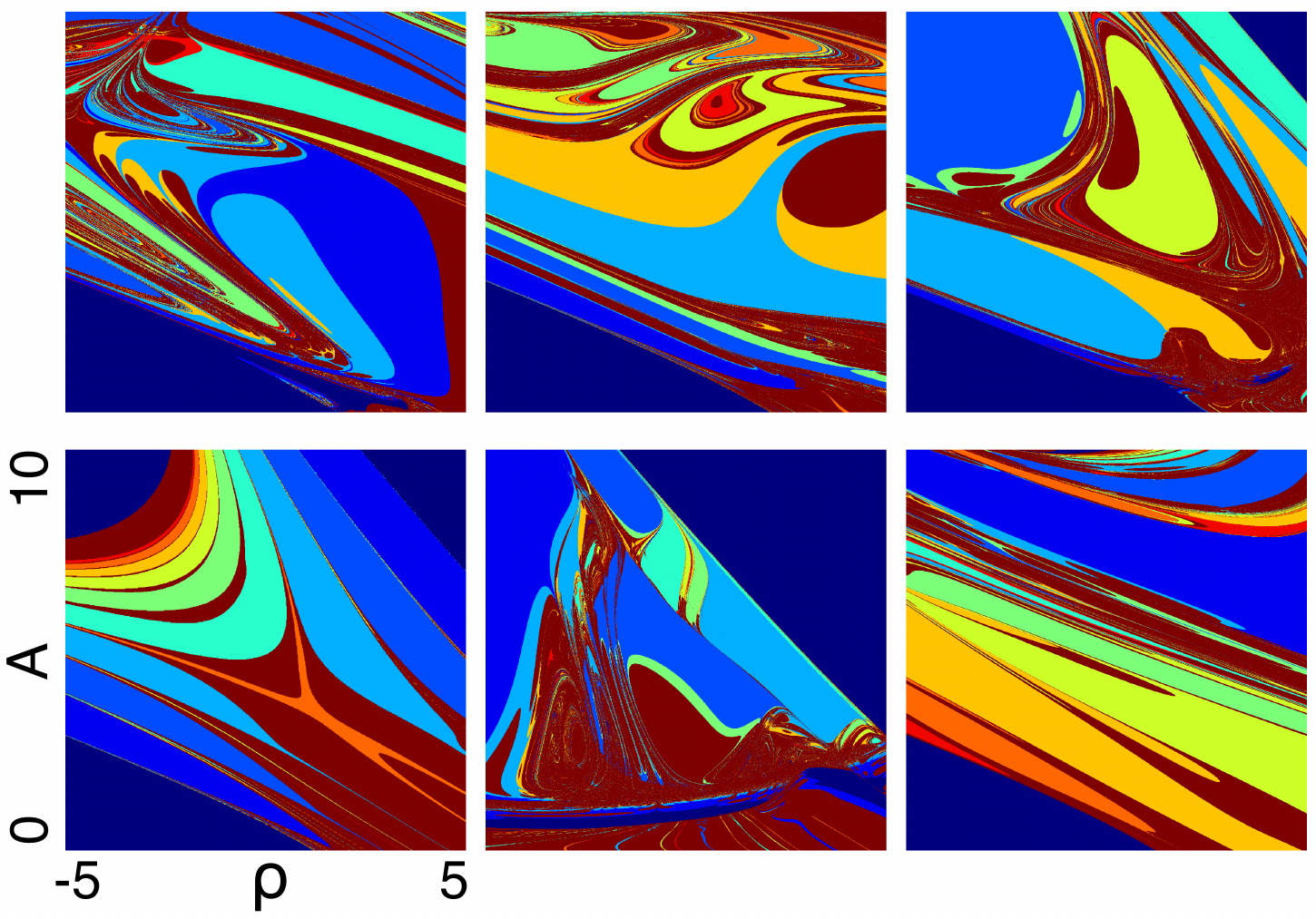}
\caption{\textbf{Stunning complexity of locking diagrams as more diversity is allowed in the circuits.} The diagrams correspond to circuits which were obtained by optimization of the objective function in a more general domain which includes the possibility of having different timescales for each population. The resulting circuits yield better scores in the optimization function and they exhibit a remarkable diversity of behaviors which are available by small modulations of the input.}
\end{figure*}

\begin{table*}
\label{parameters}
\begin{tabular}{ l c c c c c c c c }
Location & $\tau_1$ & $C_{11}$ & $C_{12}$ & $\rho_1$ & $\tau_2$ & $C_{21}$ & $C_{22}$ & $\rho_2$ \\
\hline \hline
\textbf{Fig 1.1}  &  1.00  &    5.84 & 12.84 & -16.00 &    1.00 &  -12.48 &  5.44 &  6.60 \\
Fig 1.2  &  1.00  &    15.64 &-14.32  &-0.32 &    1.00 &  19.12  & 10.64 &  -8.36 \\
Fig 1.3  &  1.00  &    1.08 & -10.72   &4.32 &    1.00 &  16.28   & 8.72 & -18.96 \\
Fig 1.4  &  1.00  &    15.56 &-9.32  &-4.80 &    1.00 & 13.08 & 14.00 &  -15.32 \\ 
Fig 1.5  &  1.00  &    8.04 & 6.80 & -10.00 & 1.00 &  -19.92 & 11.52 & -3.28 \\
\hline
\textbf{Fig 2} &  1.00  &    4.92 &  -6.76 & -3.00 &    1.00 &  14.96  & 18.76 & -14.96 \\
\hline
\textbf{Fig 3} &  1.00  &    2.32 & -17.32   &8.52 &    1.00 &  15.16 & 16.44 &-18.88 \\
\hline
\textbf{Fig 4.1.1} & 1.838 &  11.44  & -8.76  & -3.64 &   1.751 &   19.40    &10.28  &  -7.12 \\
Fig 4.1.2 & 1.8395 &  10.96  & -12.00   &   -3.68 &   1.088   &  8.4  &   10.00   &   -6.84\\
Fig 4.1.3 &  1.7705  &  14.28  &  -9.72  &  -3.64 &   0.701 &  16.96  & 7.60 & -4.92 \\
Fig 4.2.1 & 1.9145   &  14.68  &  -9.72 & -3.84 &   1.9085 &  11.32  & 10.56 & -7.80 \\
Fig 4.2.2 & 1.802 &  10.36 &  -9.44 &  0.56 &   0.545  & 16.60 & 7.24 & 5.00 \\
Fig 4.2.3 & 1.9355 &  11.20   &  -8.24 & -3.56 &   0.7745 & 12.88  & 19.28 & -16.76 \\
\end{tabular}
\caption{The parameters for all the circuits shown in this article are listed here. Circuits in Figure 1 are numbered 1 through 5 from left to right. The circuits in Figure 4 are numbered according to their row and column numbers. 
}
\end{table*}

\section{Conclusions} 

This article describes a numerical procedure designed to tune the parameters of periodically-driven dynamical systems towards regimes in which several subharmonic solutions are possible. The procedure was employed to tune the parameters of simple, low-dimensional models of neural circuits to a state in which an incoming periodic stimuli can result in many different dynamical behaviors. The procedure makes it possible to find simple architectures consisting of two populations of neurons that can accommodate such a scenario by nonlinearly entraining to the incoming signals. These circuits can be found as the result of optimizing an objective function that measures the diversity of entrainment types by computing a low resolution approximation of the locking regions of the circuits. While it is likely that a similar approach would be applicable to other systems, it is unclear whether a given system may be able to accommodate many nonlinear resonances in ranges of the parameters that are physically and biologically plausible. Neural circuits of the sort discussed in this article possess a remarkable flexibility to support this type of dynamical behavior, even in the simple case of only two interacting populations.   

Neural circuits that support an array of functions might be desirable in several contexts \cite{hoppenstead}. The procedure presented here is useful to design multipurpose pattern generators \cite{kopellcpg}. This in turn can be used as an encoder: a long complex sequence of oscillatory commands can be encoded as different responses to simpler signals consisting of parametric modulations to the circuit \cite{paramest}. The procedure presented here facilitates exploration of the hypothesis that nonlinear resonances in neural circuits may play a role in neural function by providing a tool to find the special connectivities that give rise to these dynamical properties. This procedure also enables the investigation of the dynamics of large scale models of weakly interacting neural circuits when the parameters of each circuit are tuned close to nonlinear resonances. From a dynamical systems perspective it would also be interesting to see if circuits that exhibit these properties also share other dynamical properties such as the bifurcations diagrams of the non-driven system. Finally, the procedure makes it easy to identify transitions between regimes in which universal scaling rules are expected to arise. Therefore, the topological mechanisms by which complex dynamics emerge in simple neural circuits can be systematically investigated.

\section*{Acknowledgments}
Leandro M. Alonso's research was supported by funds from a Leon Levy Fellowship at The Rockefeller University.


\begin{thebibliography}{100} 

\bibitem{buszakiscience} Buzs\'aki, Gy\"orgy, and Andreas Draguhn. \emph{Neuronal oscillations in cortical networks.} science 304, no. 5679 (2004): 1926-1929.

\bibitem{golubitsky} Golubitsky, Martin, Ian Stewart, Pietro-Luciano Buono, and J. J. Collins. \emph{Symmetry in locomotor central pattern generators and animal gaits.} Nature 401, no. 6754 (1999): 693-695.

\bibitem{delcomyn} Delcomyn, Fred. \emph{Neural basis of rhythmic behavior in animals.} Science 210, no. 4469 (1980): 492-498.

\bibitem{singernature} Engel, Andreas K., Pascal Fries, and Wolf Singer. \emph{Dynamic predictions: oscillations and synchrony in top–down processing.} Nature Reviews Neuroscience 2, no. 10 (2001): 704-716.

\bibitem{buzsakilfp} Agarwal, Gautam, Ian H. Stevenson, Antal Berényi, Kenji Mizuseki, György Buzsáki, and Friedrich T. Sommer. "Spatially distributed local fields in the hippocampus encode rat position." Science 344, no. 6184 (2014): 626-630.

\bibitem{buszakilibro} Buzsaki, Gyorgy. \emph{Rhythms of the Brain.} Oxford University Press, 2006.


\bibitem{izifm}Hoppensteadt, Frank C., and Eugene M. Izhikevich. \emph{Thalamo-cortical interactions modeled by weakly connected oscillators: could the brain use FM radio principles?.} Biosystems 48, no. 1 (1998): 85-94.

\bibitem{sejnowski} Veltz, Romain, and Terrence J. Sejnowski. \emph{Periodic forcing of inhibition-stabilized networks: Nonlinear resonances and phase-amplitude coupling.} Neural computation (2015).


\bibitem{lefebvrejon} Herrmann, Christoph S., Micah M. Murray, Silvio Ionta, Axel Hutt, and Jérémie Lefebvre. \emph{Shaping Intrinsic Neural Oscillations with Periodic Stimulation.} The Journal of Neuroscience 36, no. 19 (2016): 5328-5337.
Harvard	

\bibitem{abarbanelpyloric}Szucs, Attila, Robert C. Elson, Michail I. Rabinovich, Henry DI Abarbanel, and Allen I. Selverston. \emph{Nonlinear behavior of sinusoidally forced pyloric pacemaker neurons.} Journal of Neurophysiology 85, no. 4 (2001): 1623-1638.

\bibitem{mendez}M\'endez, Jorge M., Gabriel B. Mindlin, and Franz Goller. \emph{Interaction between telencephalic signals and respiratory dynamics in songbirds.} Journal of neurophysiology 107, no. 11 (2012): 2971-2983.

\bibitem{rabinovichroleofchaos} Rabinovich, M. I., and H. D. I. Abarbanel. \emph{The role of chaos in neural systems.} Neuroscience 87, no. 1 (1998): 5-14.

\bibitem{ottcontrol} Ott, Edward, Celso Grebogi, and James A. Yorke. \emph{Controlling chaos.} Physical review letters 64.11 (1990): 1196.

\bibitem{prlsubarmonicos} Trevisan, Marcos A., Gabriel B. Mindlin, and Franz Goller. \emph{Nonlinear model predicts diverse respiratory patterns of birdsong.} Physical review letters 96, no. 5 (2006): 058103.

\bibitem{granada} Granada, A., M. Gabitto, G. Garc\'ia, J. Alliende, J. Méndez, M. A. Trevisan, and G. B. Mindlin. \emph{The generation of respiratory rhythms in birds.} Physica A: Statistical Mechanics and its Applications 371, no. 1 (2006): 84-87.

\bibitem{pranama} Arneodo, Ezequiel M., Leandro M. Alonso, Jorge A. Alliende, and Gabriel B. Mindlin. \emph{The dynamical origin of physiological instructions used in birdsong production.} Pramana 70, no. 6 (2008): 1077-1085.

\bibitem{alonso09} Alonso, Leandro M., Jorge A. Alliende, Franz Goller, and Gabriel B. Mindlin. \emph{Low-dimensional dynamical model for the diversity of pressure patterns used in canary song.} Physical Review E 79, no. 4 (2009): 041929.

\bibitem{goldinmindlin} Goldin, Matías A., and Gabriel B. Mindlin. \emph{Evidence and control of bifurcations in a respiratory system.} Chaos: An Interdisciplinary Journal of Nonlinear Science 23, no. 4 (2013): 043138.

\bibitem{gogui}Alonso, Rodrigo G., Marcos A. Trevisan, Ana Amador, Franz Goller, and Gabriel B. Mindlin. \emph{A circular model for song motor control in Serinus canaria.} Frontiers in computational neuroscience 9 (2015).

\bibitem{gucken} Guckenheimer, John, and Philip Holmes. \emph{Nonlinear oscillations, dynamical systems, and bifurcations of vector fields.} Vol. 42. Springer Verlag: New York, 1983.

\bibitem{wiggins} Wiggins, Stephen. \emph{Introduction to applied nonlinear dynamical systems and chaos.} Vol. 2. Springer Science \& Business Media, 2003.

\bibitem{numericalrecipes} Press, William H. \emph{Numerical recipes 3rd edition: The art of scientific computing.} Cambridge university press, 2007.

\bibitem{arnold} Arnold, Vladimir. I. \emph{Geometrical Methods in the Theory of Ordinary Differential Equations (Grundlehren der mathematischen Wissenschaften).} Fundamental Principles of Mathematical Science.–Springer, Verlag, New York 250 (1983).

\bibitem{wilsoncowan72} Wilson, Hugh R., and Jack D. Cowan. \emph{Excitatory and inhibitory interactions in localized populations of model neurons.} Biophysical journal 12, no. 1 (1972): 1-24.

\bibitem{destexhewilsoncowan} Destexhe, Alain, and Terrence J. Sejnowski. \emph{The Wilson–Cowan model, 36 years later.} Biological cybernetics 101, no. 1 (2009): 1-2.

\bibitem{holland} Holland, John H. \emph{Genetic algorithms.} Scientific american 267, no. 1 (1992): 66-72.

\bibitem{feigenbaum} Feigenbaum, Mitchell J. \emph{Quantitative universality for a class of nonlinear transformations.} Journal of statistical physics 19.1 (1978): 25-52.

\bibitem{libchaber} Libchaber, A., C. Laroche, and S. Fauve. \emph{Period doubling cascade in mercury, a quantitative measurement.} Journal de Physique Lettres 43.7 (1982): 211-216.

\bibitem{ermentroutchaos} Ermentrout, G. Bard. \emph{Period doublings and possible chaos in neural models.} SIAM Journal on Applied Mathematics 44, no. 1 (1984): 80-95.

\bibitem{alicestretchland} Gilmore, Robert, and Marc Lefranc. \emph{The topology of chaos: Alice in stretch and squeezeland}. John Wiley \& Sons, (2008).

\bibitem{classification} Mindlin, Gabriel B., Xin-Jun Hou, Hernán G. Solari, R. Gilmore, and N. B. Tufillaro. \emph{Classification of strange attractors by integers.} Physical review letters 64, no. 20 (1990): 2350.

\bibitem{yorke} Kennedy, Judy, and James A. Yorke. \emph{Basins of Wada.} Physica D: Nonlinear Phenomena 51, no. 1-3 (1991): 213-225.

\bibitem{hoppenstead} Hoppenstead F. C. (Editor). \emph{Nonlinear oscillations in biology.} Lectures in Applied Mathematics Volume 17. American Mathematical Society, (1979).

\bibitem{kopellcpg} Kopell, Nancy, and G. Bard Ermentrout. \emph{Coupled oscillators and the design of central pattern generators.} Mathematical biosciences 90, no. 1 (1988): 87-109.

\bibitem{paramest} Alonso, Leandro M. \emph{Parameter estimation, nonlinearity, and Occam's razor.} Chaos: An Interdisciplinary Journal of Nonlinear Science 25, no. 3 (2015): 033104.


\end{thebibliography}
\end{document}